\documentclass[aps,prb,twocolumn,floats,showpacs]{revtex4}
\usepackage[dvips]{graphicx} \usepackage{epsfig} \usepackage{amsmath}
\usepackage{times} \usepackage{amssymb} \usepackage{amsbsy}
\begin{document}
\title{Kondo-lattice model: Application to the temperature-dependent
  electronic structure of EuO(100) films } 
\author{R.~Schiller}
\author{W.~M\"uller}
\email[Email: ]{Wolf.Mueller@physik.hu-berlin.de}
\homepage{http://tfk.physik.hu-berlin.de}
\author{W.~Nolting}
\affiliation{Lehrstuhl Festk\"orpertheorie, Institut f\"ur Physik,
  Humboldt-Universit\"at zu Berlin,\\ Invalidenstr.\ 110, 10115 Berlin,
  Germany}
\date{\today}
%\maketitle
\begin{abstract}
  We present calculations for the temperature-dependent electronic
  structure and magnetic properties of thin ferromagnetic EuO films. The
  treatment is based on a combination of a multiband-Kondo lattice model
  with first-principles TB-LMTO band structure calculations. The method
  avoids the problem of double-counting of relevant interactions and
  takes into account the correct symmetry of the atomic orbitals. We
  discuss the temperature-dependent electronic structures of EuO(100)
  films in terms of quasiparticle densities of states and quasiparticle
  band structures. The Curie temperature $T_{\mathrm{C}}$ of the EuO films turns
  out to be strongly thickness-dependent, starting from a very low value
  ($\simeq 15K$) for the monolayer and reaching the bulk value at about
  25 layers.  \index{abstract}
\end{abstract}

\maketitle
\section{Introduction}\label{sec:introduction}
A highly vivid field of research in current solid state physics aims at
the intercorrelation between electronic and magnetic properties of
materials. Special highlights in this respect include phenomena like the
colossal magnetoresistance (CMR) effect observed in the manganites
perowskites $T_{1-x}D_{x}MnO_3$ where T is a trivalent lanthanide ion
such as La and D is a divalent alkaline-earth (e.g.\ Ca, Ba, Sr) ion
\cite{Ram97}. The interest is fueled by the prospect for applications in
the fields of storage media and sensor technology.  Interesting from an
even more fundamental point of view is the intercorrelation of
electronic and magnetic properties in the so-called `local-moment`
magnets, prototypically realized by the metallic rare earths Gd, Tb and
Dy \cite{Ley80} and the insulating rare-earth compounds EuO and EuS
\cite{Wac79}. Sometimes these materials are also denoted as `4f systems`
because a great part of their properties is due to the existence of a
partially filled and highly localized 4f-shell belonging to the
rare-earth ion. Strong screening by other closed electron shells keeps
the Hund's rules valid. According to these the 4f electrons couple to a
finite moment which is strictly localized at the sites of the rare-earth
ion ($Eu^{2+}, Gd^{3+}, ...$). While the purely magnetic properties of
these materials are due to their localized `magnetic` 4f electrons the
conductivity properties are determined by the quasi-free electrons in
rather broad (5d, 6s) conduction bands.

Many characteristic features of 4f systems follow from an intimate
correlation between localized and itinerant electrons. This correlation
leads, e.g., to a striking temperature dependence of the conduction band
states induced by the magnetization state of the local moment system.
First evidence for this has been found in 1964 by the discovery of the
`red shift effect` in EuO by Busch and Wachter \cite{Wac79}. The effect
refers to a strong shift of the lower band edge to lower energies upon
temperature cooling below $T_{\mathrm{C}}$. A further striking effect being due to
the mentioned induced temperature dependence is a metal-insulator
transition observed in Eu-rich EuO \cite{Pen72}. The Eu-richness
manifests itself in (2+)-charged oxygen vacancies. One of the two
excesses electrons is thought to be tightly bound to the vacancy while,
because of the Coulomb repulsion, the other occupies an impurity level
fairly close to the lower band edge. When cooling below $T_{\mathrm{C}}$ the band
edge crosses the impurity level (`red shift`) freeing therewith the
excess electrons. A conductivity jump of up to 14 orders of magnitude
has been observed.

Further effects result from the interaction of the band electron with
collective excitations of the moment system. One of these is the
creation of a new quasiparticle which is called `magnetic polaron`. It
can be classified as a propagating electron dressed by a virtual cloud
of excited magnons. The formation of magnetic polarons strongly
influences the electronic structure of respective materials
\cite{NMR96,NMR97,SN99}.

The influence between localized and itinerant electrons is of course of
mutual character. That means that a finite band occupation has
observable consequences for the moment system, too. The ferromagnetic
order in Gd, e.g., can be explained only by indirect exchange coupling
(RKKY) of the localized moments (spins) mediated by a spin polarization
of the a priori non-magnetic conduction electrons. This RKKY-mechanism
can excellently be monitored by inspecting the alloy $Eu_{1-x}Gd_{x}S$.
Replacing the $Eu^{2+}$-ion in the ferromagnetic insulator EuS by a
$Gd^{3+}$ leads in a definite manner to a population of the conduction
band without diluting the local moment system. The latter holds because
$Eu^{2+}$ and $Gd^{3+}$ both have a half-filled 4f-shell
$S=\frac{7}{2}$. That allows for a direct study of the carrier
concentration dependence of the Curie temperature
$T_{\mathrm{C}}=T_{\mathrm{C}}(n)$\cite{SW??,GP80}.

It is commonly accepted that the so called s-f model \cite{Zen51,Nol79}
provides a good starting point for an at least qualitative understanding
of the above properties of typical `local moment` materials. Within the
model they are traced back to an intraatomic, interband exchange
interaction between localized and extended states. The same model has
been extensively applied in the recent past to the CMR-materials. In the
manganites the three almost localized $3d_{t_{2g}}$-electrons couple to
an $S=\frac{3}{2}$-spin that is exchange-coupled to the somewhat more
mobile $3d_{e_g}$-electrons. In this context the s-f model is referred
to as the ferromagnetic Kondo-lattice model (FKLM)
\cite{Fur94,Fur98,DYM98}. The additive `ferromagnetic` points to an
exchange coupling which favors the parallel alignment of itinerant and
localized spin. Without this additive the `normal` Kondo-lattice model
(KLM) is meant that differs from the FKLM only by a sign change of the
coupling constant favoring therewith an antiparallel alignment. The
latter version is considered a candidate for the extraordinary
heavy-fermion physics \cite{GS91}. In this paper we exclusively refer to
the FKLM, which covers the above described 4f-systems and the
manganites. While the 4f-systems surely belong to the weak-coupling
region (exchange coupling much smaller than the bandwidth) the
manganites are to be ascribed to the strong coupling limit.

A central aspect of our study which we are going to present in this
paper is that of reduced dimensionality and its interplay with typical
correlation effects. Magnetic phase transitions and electronic
structures at surfaces and in thin films have attracted a lot of recent
research work. One of the most remarkable magnetic phenomena observed in
4f-systems is the existence of a ferromagnetic surface at temperatures
where the bulk material is already paramagnetic. This effect was first
documented for the Gd(0001) surface by Weller et al. \cite{RE86} and
since then has been measured by several groups with different
experimental techniques. Depending on the used techniques and maybe on
the cleanness of the applied surface Curie-temperature enhancements
$\Delta T_{C}=T_{\mathrm{C}}(\mathrm{surface})-T_{\mathrm{C}}(\mathrm{bulk})$ in between
17K and 60K have 
been found \cite{RE86,WAG85,TWW93}. 
%%% begin inserted
However, the possibility of an enhanced surface Curie-temperature
remains a matter of controversy. Donath et al.\ \cite{DGP96} using
spin-resolved photoemission did not find any indication for $\Delta
T_{\mathrm{C}}>0$ at Gd(0001) surfaces. By careful analysis applying
spin-polarized secondary-electron emission spectroscopy to surface
magnetization measurements and magneto-optic Kerr effect to bulk
magnetization measurements Arnold and Pappas \cite{AP00} come to the
same conclusion.
%%% end inserted
%%% begin changed
With respect to a possible $\Delta T_{\mathrm{C}}$ at Gd
%%% end changed
(also Tb) surfaces the existence of a Gd(0001) surface state seems to
play a crucial role. Predicted by band structure calculations
\cite{WF91} and simultaneously measured with photoemission
\cite{LHD91,FSK94,WSM96} in particular its temperature dependent induced
exchange splitting has attracted attention \cite{DDN98}. Is it
`Stoner-like` collapsing for $T \longrightarrow T_{\mathrm{C}}$ or does the
splitting persist in the paramagnetic phase performing the
demagnetization by a redistribution of spectral weight (`spin-mixing
behavior`) \cite{WSM96,DDN98}? It is an open question whether and how a
surface state can cause an enhanced surface-Curie temperature. To get
%%% begin changed
insight a calculation of the temperature-dependent electronic structure
%%% end changed
of the real (!) Gd-film would be required. For providing the clues to
such a treatment of local-moment metals we present in this paper the
methodical approach for dealing with local-moment semiconductor films,
which are in addition interesting in themselves. The semiconducting
4f-systems EuO, EuS may be considered as `low-density limits` of the
metallic Gd. They have the same half-filled 4f-shell as Gd producing
localized $S=\frac{7}{2}$-spins. They differ by the carrier
concentration n ($n=0$ in EuO, EuS; $n\neq 0$ in Gd). We present in this
paper the description of a real EuO-film of variable thickness. We are
interested in the layer-dependent magnetic and electronic properties.
For this purpose we combine a many-body treatment of the FKLM with a
`first-principle` electronic structure calculation. For the many-body
part we use a `moment-conserving decoupling procedure` (MCDA) the idea
of which is developed in ref. 6 for the single-band bulk-FKLM. The
reformulation for the special situation of a film geometry has been
performed in \cite{SN99}. We present here the generalization to a
multi-band-FKLM film. As to the real EuO (100) film we combine the model
calculation with a `tight-binding linear muffin tin orbital` (TB-LMTO)
band structure calculation to get temperature- and layer-dependent
magnetizations, quasiparticle densities of states (Q-DOS) and
quasiparticle band structures (Q-BS). We find for thicker films ($\geq
20$ monolayers) a surface state. Being of course unoccupied it
nevertheless exhibits a temperature-dependent exchange splitting induced
by exchange coupling to the magnetic 4f-moment system. We speculate that
this surface state can be considered an analogous to the Gd-surface state
mentioned above. Furthermore, we find some evidence that the temperature
behavior of the EuO-surface state together with the red shift of the
lower conduction band edge may close the 4f-5d gap in the $
\uparrow$-spectrum giving rise to a
surface-insulator-halfmetal-transition.  
%%% begin deleted
%The paper is organized as
%follows: In the next section we introduce the multiband-Kondo lattice
%Hamiltonian and outline how to perform the ``first principles''
%bandstructure calculation for the EuO--film. In
%Sect.~\ref{sec:many-body-evaluation} the procedure sketched which we
%have used to solve (approximately) the many-body problem of the
%multiband Kondo lattice. Du to the zero band occupation, the problem can
%be divided into an electronic problem and a magnetic problem, which can
%be treated separately. The magnetic problem is that of an effective
%Heisenberg model. A Green--function technique is used to approach the
%electronic part, when the effective, in principle all interaction
%incorporating $T=0$--results of the bandstructure calculation are
%implemented in the single--electron part of the multiband-FKLM
%Hamiltonian. Results concerning the electronic and magnetic properties
%of EuO (100) films are presented in Sect.~\ref{sec:results-euo-films}.
%%% end deleted

\section{Multiband-Kondo lattice model}\label{sec:mult-kondo-latt}
\subsection{Model-Hamiltonian}\label{sec:model-hamiltonian}
We investigate a film consisting of $n$ equivalent layers parallel to
the surface of the film. Each lattice site of the film is indicated by a
Greek letter $\alpha$, $\beta$, $\gamma$, $\ldots$ denoting the layer
index and a Latin letter $i$, $j$, $k$, $\ldots$ numbering the sites
within a given layer.  Each layer possesses two-dimensional
translational symmetry. Accordingly, the thermodynamic average of any
site dependent operator $A_{i\alpha}$ depends only on the layer index
$\alpha$:
\begin{equation}
\label{eq:site_ind_ev}
\left\langle A_{i\alpha} \right\rangle \equiv \left\langle 
A_{\alpha} \right\rangle.
\end{equation} 
The complete {\em d-f} model Hamiltonian for a real system with multiple
conduction bands,
\begin{equation}
\label{eq:sf_ham}
{\cal H} = {\cal H}_d + {\cal H}_f + {\cal H}_{df} ,
\end{equation}
consists of three parts. The first
\begin{equation}
  \label{eq:h_d}
  {\cal H}_d = \sum_{ij \alpha\beta\sigma} \sum_{mm'}
  T^{mm'}_{ij\alpha\beta} c^+_{i\alpha m\sigma} c_{j\beta m'\sigma}
\end{equation}
contains the conduction band structure of e.g.\ EuO. The indices $m$ and
$m'$ denote the different conduction bands.  $c^+_{i\alpha m\sigma}$ and
$c_{i\alpha m\sigma}$ are, respectively the creation and annihilation
operators of a spin-$\sigma$ electron from the $m$'th subband at the
lattice site ${\bf R}_{i\alpha}$.  $T^{mm'}_{ij\alpha\beta}$ are the
hopping integrals which later have to be obtained from an LDA
calculation.

Each lattice site ${\bf R}_{i\alpha}$ is occupied by a localized
magnetic moment, represented by a spin operator ${\bf S}_{i\alpha}$. In
the case of Eu and Gd the {\em 4f} shell is exactly half-filled and,
because of Hund's rules, has its maximal magnetic moment of $7\,\mu_{\rm
  B}$.  These localized moments are described by an extended Heisenberg
Hamiltonian
\begin{equation}
\label{eq:h_f}
{\cal H}_f = \sum_{ij\alpha\beta} J^{\alpha\beta}_{ij} 
{\bf S}_{i\alpha} {\bf S}_{j\beta} + D_0
  \sum_{i\alpha} (S^z_{i\alpha})^2.
\end{equation}
The first term is the original Heisenberg Hamiltonian with the exchange
integrals $J^{\alpha\beta}_{ij}$.  The second term introduces a
symmetry-breaking single-ion anisotropy, which is necessary for film
geometries to obtain a collective magnetic order at finite temperatures
\cite{MW66,GN00}. Here, one has typically $D_0 \ll
J^{\alpha\beta}_{ij}$.

The distinguishing feature of the single-band Kondo-lattice model (s-f
or d-f model) is an intraatomic exchange between the conduction
electrons and the localized f-spins. The form of the respective
Hamiltonian in the case of multiple conduction bands can be derived from
the general form of the on-site Coulomb interaction between electrons of
different subbands.
\begin{equation}
{\cal H}_{I}=\frac{1}{2} \sum_{L_{1}\ldots L_{4}} 
\sum_{\sigma\sigma^{\prime}}U_{L_{1}\ldots L_{4}}c^{+}_{L_{1}\sigma}
c^{+}_{L_{2}\sigma^{\prime}}c_{L_{3}\sigma^{\prime}}c_{L_{4}\sigma}
\end{equation}
We drop for the moment the site index i because we want to keep the
local character of the s-f interaction. $L_1 \ldots L_4$ denote the
different bands, and $U_{L_1 \ldots L_4}$ are the Coulomb matrix
elements. Restricting the electron scattering processes caused by the
Coulomb interaction to two involved subbands we get instead of
(\ref{eq:h_i}): 
\begin{eqnarray}
  \label{eq:h_i}
{\cal H}_I=\frac{1}{2}
\sum_{L,L^{\prime}}\sum_{\sigma\sigma^{\prime}}[U_{L,L^{\prime}}c^+_{L\sigma}c^+_{L^{\prime}\sigma^{\prime}}c_{L^{\prime}\sigma^{\prime}}c_{L\sigma}+\nonumber\\
J_{L,L^{\prime}}c^+_{L\sigma}c^+_{L^{\prime}\sigma^{\prime}}c_{L\sigma^{\prime}}c_{L^{\prime}\sigma}+
\nonumber\\J^*_{L,L^{\prime}}c^+_{L\sigma}c^+_{L\sigma^{\prime}}c_{L^{\prime}\sigma^{\prime}}c_{L^{\prime}\sigma}]
\end{eqnarray}
Thinking of 4f systems such as EuO and Gd the band indices L and L' can
be attributed either to a flat 4f band ($L\longrightarrow f$) or to a
broad (5d,6s) conduction band ($L \longrightarrow m$). In an obvious
manner we can then split the Coulomb interaction into three different
parts
\begin{equation}
{\cal H}_I={\cal H}^{cc}_I+{\cal H}^{ff}_I+{\cal H}^{cf}_I
\end{equation}
depending on whether both interacting particles stem from a conduction
band, ${\cal H}^{cc}_I$, or both from a flat band,${\cal H}^{ff}_I$, or
one from a flat band the other from a conduction band,${\cal H}^{cf}_I$.
The first term refers to electron correlations in the conduction bands,
which by definition is neglected in the KLM. In the case of EuO this
simplification becomes exact because the conduction bands are unoccupied
($n=0$). On the other hand, for the CMR-materials the neglect of ${\cal
  H}^{cc}_I$ appears questionable. Some authors therefore extend the KLM
to the `correlated` KLM by regarding ${\cal H}^{cc}_I$ as the
Hubbard-interaction \cite{HV00}. The second term,${\cal H}^{ff}_I$, is
already contained in the description of the localized moments via the
Heisenberg model~(\ref{eq:h_f}). So we are left with the interaction between
localized and itinerant electrons:
\begin{eqnarray}
{\cal H}^{cf}_I=\sum_{m,f,\sigma ,\sigma^{\prime}}[U_{mf}c^+_{m\sigma}c^+_{f\sigma^{\prime}}c_{f\sigma^{\prime}}c_{m\sigma}+\nonumber\\
J_{mf}c^+_{m\sigma}c^+_{f\sigma^{\prime}}c_{m\sigma^{\prime}}c_{f\sigma}+
\nonumber\\\frac{1}{2}J^*_{mf}c^+_{m\sigma}c^+_{m\sigma^{\prime}}c_{f\sigma^{\prime}}c_{f\sigma}+\nonumber\\\frac{1}{2}J^*_{fm}c^+_{f\sigma}c^+_{f\sigma^{\prime}}c_{m\sigma^{\prime}}c_{m\sigma}]
\end{eqnarray}
The last two terms do not contribute in case of the 4f systems (EuO,Gd)
since the ($Eu^{2+},Gd^{3+}$)-4f shell has its maximum spin of $
S=\frac{7}{2}$. All the seven 4f electrons have to occupy different
subbands and none of the 7 sublevels will be doubly occupied.
Introducing electron spin operators,
\begin{eqnarray}
\sigma^+ &=& \hbar\, c^+_{\uparrow}c_{\downarrow}\nonumber\\
\sigma^- &=& \hbar\, c^+_{\downarrow}c_{\uparrow}\nonumber\\
\sigma^z &=& \frac{\hbar}{2}\, (n_{\uparrow}-n_{\downarrow}) 
\end{eqnarray}
one arrives at
\begin{eqnarray}
{\cal H}^{cf}_I&=&-\frac{2}{\hbar^2}\sum_{mf}J_{mf}\sigma_m\sigma_f + \nonumber\\
&&+\sum_{mf}(U_{mf}-\frac{1}{2}J_{mf})\,n_m\, n_f
\end{eqnarray}
with $n_{m(f)}=n_{m(f)\uparrow}-n_{m(f)\downarrow}$. In the case of EuO,
we are interested in, the last term does not contribute because of $n_m
=0$. Even for Gd ($n_m\neq 0$) it certainly can be neglected since the
number of electrons is fixed, $n_f$ thus being only a c-number. By
defining the spin operator ${\bf S}$ of the local f-moment,
\begin{equation}
{\bf S}=\sum_f \sigma_f
\end{equation}
and by assuming the interband exchange $J_{mf}$ to be independent on the
band indices m and f,
\begin{equation}
J_{mf}\equiv -\frac{1}{2}\hbar J
\end{equation}
the interaction term eventually reads,
\begin{equation}
{\cal H}^f_I=-\frac{J}{\hbar}\sum_m\,\sigma_m {\bf S}
\end{equation}
Reintroduction of the lattice site dependence ($R_{i\alpha}$) we end up
with the interaction term of the multiband-KLM:
\begin{eqnarray}
\label{eq:H_df}
{\cal H}_{d f}&=&-\frac{J}{\hbar}\sum_{i\alpha m}\sigma_{i\alpha m}\,{\bf S}_{i\alpha}\nonumber\\
&=&-\frac{1}{2} J\sum_{i\alpha m\sigma}[z_{\sigma}\,{\bf S}^z_{i\alpha}+\nonumber\\
&&+{\bf S}^{\sigma}_{i\alpha m }\,c^+_{i\alpha m -\sigma}\,c_{i\alpha m \sigma}]
\end{eqnarray}
Here we have used the abbreviations
\begin{eqnarray}
{\bf S}_{j\alpha m}^{\sigma} = {\bf S}_{j\alpha m}^x + i\,z_{\sigma}\,{\bf S}_{j\alpha m}^y\nonumber\\z_{\sigma}=\delta_{\sigma\uparrow}-\delta_{\sigma\downarrow}
\end{eqnarray}
Compared to the conventional single-band KLM we have simply an
additional band-summation in the exchange term. Note, however, the
subbands are intercorrelated via the single-particle term ${\cal H}_d$
(\ref{eq:h_d}).
\subsection{Band structure calculations}
The hopping integrals $T^{mm^{\prime}}_{ij\alpha\beta}$ in the
%%% begin changed
single-particle Hamiltonian (\ref{eq:h_d}) has to contain the influences
of all
%%% end changed
those interactions which are not directly covered by our
model-Hamiltonian. For this reason we perform a band structure
calculation according to the `tight binding`-LMTO-ASA program of
Anderson \cite{And75,AJ84}. The resulting single-particle energies are
taken as single-particle input for the partial Hamiltonian (\ref{eq:h_d}).

Test calculations for bulk-EuO have revealed the well known `gap
problem` which appears in connection with the strongly localized
character of the 4f-'bands'. A `normal` LDA calculation makes EuO a
metal with the 4f-levels lying well within the conduction band. To
overcome this situation we have investigated two approaches for dealing
with the localized 4f moments (see \cite{EAO95} for the analogous case
of Gd). In the first, the 4f moments are treated as localized core
electrons, which naturally gives an insulator ground state, while the
conduction band region consists predominantly of Eu-5d states. In the
second, an LDA+U calculation \cite{AZA91,AAL97} has been performed,
which has the advantage compared to the former procedure, that the flat
4f bands explicitly appear right between the O-2p bands and the Eu-5d
bands. In this approach, a meanfield approximation for the direct
Coulomb U and the exchange interaction between the 4f electrons is added
to the LDA-energy functional. The parameters U and J can be used to
position the 4f levels correctly in the band gap between the O-2p and
the Eu-5d levels. Besides the fact, that adjustable parameters U and J
somehow corrupt the idea of a `first principles` calculation, and since
we consider the 4f-levels in our model study only as localized magnetic
moments (\ref{eq:h_f}), the much simpler LDA calculation has to be preferred in
particular when our study focuses on more complicated film geometries.
\begin{figure}[t]
  \centerline{\epsfig{file=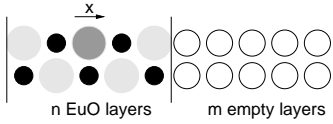,width=0.5\linewidth}}
    \caption{Supercell geometry for the EuO(100) film calculations, with
      the Europium atoms (large), the oxygen atoms (small) and the empty
      spheres (circles). The vertical lines indicate the surface of the
      EuO film.}
    \label{fig:Nol:fig1}
\end{figure}
For the band-structure calculations- for EuO films one has to employ a
supercell geometry as depicted in Fig.~\ref{fig:Nol:fig1}. As a result,
the system has a super-layered structure with super-layers consisting of
n consecutive EuO(100) layers followed by m layers of empty spheres.
That means, we have a system of periodically stacked EuO n-layer films,
isolated from each other by m layers of empty spheres.

\begin{figure}[htp]
  \centerline{\epsfig{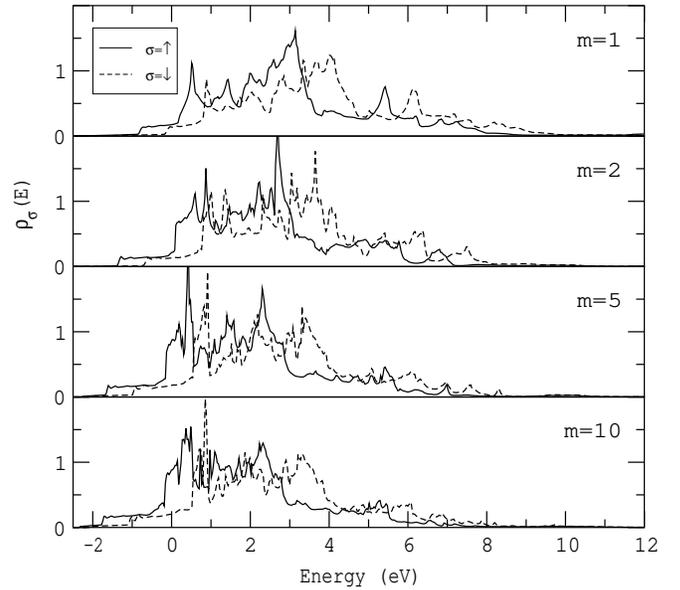}}
    \caption{Partial densities of states for the Eu-5d bands of a
      EuO(100) monolayer calculated for different numbers m of spacer
      layers of empty spheres. %Top: Bulk EuO for comparison.
      }
    \label{fig:Nol:fig2}
\end{figure}

\begin{figure}[t]
  \centerline{\epsfig{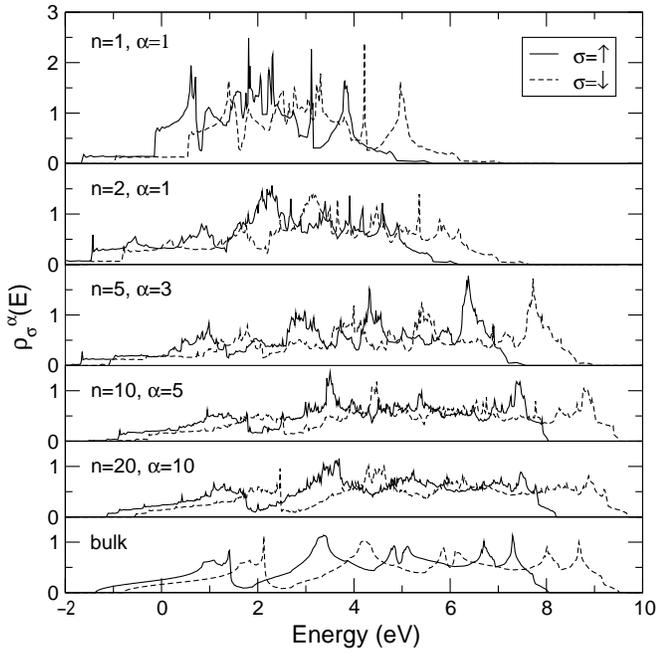}}
    \caption{Partial densities of states for the Eu-5d bands of the
      center layers ($\alpha$: layer index) for EuO films of various
      thicknesses n. Bottom: Bulk EuO}
    \label{fig:Nol:fig3}
\end{figure}

The number m of empty layers has to be chosen large enough to have truly
isolated EuO films, but on the other hand also as small as possible to
cut down on the numerical effort. In Fig.~\ref{fig:Nol:fig2}, the
5d-partial density of states of EuO(100) monolayers (n=1) is displayed
for different numbers m of spacer layers. The DOS of the EuO monolayer
converges quite quickly as a function of m. For the following we have
chosen m=5 which, according to Fig.~\ref{fig:Nol:fig2}, seems to be a
reasonable approximation for the case of ideally isolated EuO films
($m\longrightarrow \infty$).
Fig.~\ref{fig:Nol:fig3}. shows the partial densities of states of the 5d
bands of the center layer of films of various thicknesses and for bulk
EuO. The respective curves converge as a function of film thickness.
Comparing the partial Eu-5d DOS of the center layer ($\alpha =10$) of a
20-layer EuO(100) film with that of bulk EuO, we see that apart from an
unimportant constant energy shift the main features of the densities of
states match. Consequently, the center layer at the 20-layer film can be
regarded as a bulk-like environment. The still existing slight
discrepancy may be accounted for by the fact that the supercell geometry
in the band structure calculations for the EuO films destroys the cubic
symmetry present in bulk EuO. Therefore, the splitting of the d-states
into $t_{2g}$ and $e_g$ orbitals is not valid anymore.
Furthermore, the center of a 20-layer EuO(100) film of course can be
only an approximation in bulk EuO. Thus we believe that the slight
discrepancy between the partial Eu-5d densities of states of the center
layer of the 20-layer film and of bulk EuO is acceptable.

\begin{figure}[t]
  \centerline{\epsfig{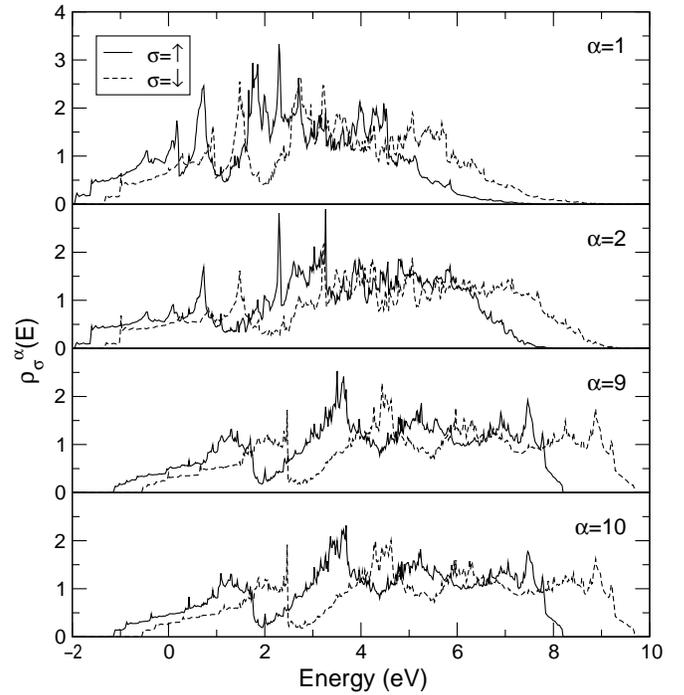}}
    \caption{Partial densities of states for the Eu-5d bands for
      different layers $\alpha$ of a 20-layer EuO(100) film.}
    \label{fig:Nol:fig4}
\end{figure}

In Fig.~\ref{fig:Nol:fig4} the partial Eu-5d densities of states of a
20-layer film are displayed for the two surface layers, $\alpha =1,2$,
and for the two center layers, $\alpha =9,10$. We see that the local
density of states in the center of the film is constant, but different
from that in the vicinity of the surface of the film. The lower band
edge of the Eu-5d bands of the surface layer compared to those of the
center layer, which represent a bulk-like situation, indicates the
existence of surface states \cite{SN01}. This interesting point will be
discussed a little bit later.
When we take the results of the just-described band structure
calculation as input for our finite temperature quasiparticle band
structure calculation, then we are running into the well known `double
counting problem` of just the relevant interactions, once explicitly in
the model-Hamiltonian and besides that implicitly in the above band
structure calculation. How we avoid this problem is explained in the
next Section.

\section{Many-body evaluation}\label{sec:many-body-evaluation}
After having fixed the model-Hamiltonian (\ref{eq:sf_ham}) and its renormalized
single-particle energies in (\ref{eq:h_d}) via a full band structure calculation we
have now to solve the many-body problem provoked by the Hamiltonian H.
There are two partial problems to be solved, one concerning the magnetic
properties of the localized 4f moments , the other dealing with the
temperature-reaction of the conduction band states on the magnetic state
of the moment system. Since we are interested in the description of the
ferromagnetic semiconductor EuO, we can assume an empty conduction band
(n=0), more strictly, a single electron in otherwise empty conduction
bands. Therefore, the magnetic order cannot be influenced by conduction
electrons and the magnetic part can be solved separately.

\subsection{Local moment system}\label{sec:local-moment-system}
The system of localized {\em f}-moments is described by the extended
Heisenberg Hamiltonian (\ref{eq:h_f}).  Here we want to stress once more
that the single-ion anisotropy constant $D_0$ is small compared to the
Heisenberg exchange interaction, $D_0 \ll J^{\alpha\beta}_{ij}$. By
defining the magnon Green function
\begin{equation}
  \label{eq:green_heis}
  P^{\alpha\beta}_{ij}(E) = \left\langle\!\left\langle
  S^+_{i\alpha}; S^-_{j\beta} \right\rangle\!\right\rangle_E,
\end{equation}
we can calculate the {\em f}-spin correlation functions by evaluating
the equation of motion
\begin{eqnarray}
  \label{eq:eom_green_heis}
  E\,P^{\alpha\beta}_{ij}(E)&=& 2 \hslash^2 \delta_{\alpha\beta}
  \langle S^z_{\alpha} \rangle \nonumber\\
 && + \left\langle\!\left\langle \left[
        S^+_{i\alpha}, {\cal H}_f \right]_- ; S^-_{j\beta}
        \right\rangle\!\right\rangle_E .
\end{eqnarray}
The evaluation of this equation of motion involves the decoupling of the
higher Green functions on its right-hand side, originating from the
Heisenberg exchange term, and the anisotropy term using the Random Phase
Approximation (RPA) and a decoupling proposed by Lines \cite{Lin67},
respectively.  The details of the calculation can be found in a previous
paper \cite{SN99b}. For brevity we restrict ourselves to present here
only the results.  For the layer-dependent magnetizations of the {\em
  f}-spin system we get\cite{Cal63}
\begin{equation}
  \label{eq:s_z}
  \frac{\langle S^z_{\alpha}\rangle}{\hslash}  = 
  \frac{(1+\varphi_{\alpha})^{2S+1} (S-\varphi_{\alpha}) +
    \varphi^{2S+1}_{\alpha} (S+1+\varphi_{\alpha})}
    {\varphi^{2S+1}_{\alpha} - (1+\varphi_{\alpha})^{2S+1}},
\end{equation}
where
\begin{equation}
  \label{eq:varphi}
  \varphi_{\alpha} = \frac{1}{N} \sum_{{\bf k}} \sum_{\gamma}
  \frac{\chi_{\alpha\alpha\gamma}({\bf k})}
  {{\rm e}^{\beta E_{\gamma}({\bf k})} - 1},
\end{equation}
where, again, $N$ is the number of sites per layer and
$\beta=\frac{1}{k_{{\rm B}}T}$.  The summation $\sum_{\gamma}$ in Eq.\ 
(\ref{eq:varphi}) runs over the $n$ poles $E_{\gamma}({\bf k})$ of the
Green function $P^{\alpha\beta}_{{\bf k}}(E)$ and the
$\chi_{\alpha\alpha\gamma}({\bf k})$ is the weight of the $\gamma$'th
pole in the diagonal element of the Green function
$P^{\alpha\alpha}_{{\bf k}}(E)$. The poles and the weights can be
calculated from the RPA solution of Eq.\ (\ref{eq:eom_green_heis}):
\begin{equation}
  \label{eq:sol_eom_heis}
  P^{\alpha\beta}_{{\bf k}}(E) = 2 \hslash^2
  \left( \begin{array}{ccc}\langle S^z_1 \rangle & & 0 \\
  & \ddots & \\ 0 & & \langle S^z_n \rangle \end{array} \right)
  \cdot (E\, {\bf I} - {\bf A})^{-1},
\end{equation}
with
\begin{equation}
  \label{eq:a_al_be}
  \frac{({\bf A})^{\alpha\beta}}{\hslash} = \big( D_0 \Phi_{\alpha} +
  2 \sum_{\gamma} J^{\alpha\gamma}_{{\bf 0}} \langle S^z_{\gamma}
  \rangle \big) \delta_{\alpha\beta} - 2 J^{\alpha\beta}_{{\bf k}}
  \langle S^z_{\alpha} \rangle.
\end{equation}
The $\Phi_{\alpha}$ come from the decoupling of the higher Green
function on the right-hand side of Eq.\ (\ref{eq:eom_green_heis}) and
originates from the anisotropy term in the Hamiltonian ${\cal H}_{f}$
according to Lines \cite{Lin67}:
\begin{equation}
  \label{eq:lines}
  \Phi_{\alpha} = \frac{2\hslash^2 S(S+1) - 3 \hslash \langle
    S^z_{\alpha} \rangle (1+2 \varphi_{\alpha})}{\langle S^z_{\alpha} \rangle}.
\end{equation}

Having obtained the layer-dependent magnetizations (\ref{eq:s_z}) and
the $\varphi_{\alpha}$ we can now express all the other {\em f}-spin
correlation functions which we need for the following via the relations
\begin{subequations}
  \label{eq:rel_coeff}
  \begin{eqnarray}
    \label{eq:rel_coeff_s-s+}
    \langle S^-_{\alpha} S^+_{\alpha} \rangle & = & 
    2 \hslash \langle S^z_{\alpha} \rangle \varphi_{\alpha}, \\
    \langle (S^z_{\alpha})^2 \rangle & = & \hslash^2 S (S+1)
    \langle S^z_{\alpha} \rangle (1+2\varphi_{\alpha}),\\
    \langle (S^z_{\alpha})^3 \rangle & = & \hslash^3 S(S+1)
    \varphi_{\alpha}\nonumber\\ &&+ \hslash^2 \langle S^z_{\alpha} \rangle \big(
    S(S+1) + \varphi_{\alpha} \big) \nonumber \\
    & & - \hslash \langle (S^z_{\alpha})^2 \rangle (1+2\varphi_{\alpha}),
  \end{eqnarray}
\end{subequations}
and the general spin-operator equality
\begin{equation}
  \label{eq:spin_op_rel}
  S^{\sigma}_{i\alpha} S^{-\sigma}_{i\alpha} = \hslash^2 S(S+1) +
  z_{\sigma} \hslash S^z_{i\alpha} - (S^z_{i\alpha})^2 .
\end{equation}

We shall see in the next Section that these {\em f}-spin correlation
functions provide the whole temperature dependence of the electronic
subsystem (\ref{eq:h_*}).
\subsection{Conduction bands}\label{sec:conduction-bands}
Similarly as the magnetic part in the preceding Section the electronic
part can be treated separately in the case of EuO. The reason is that
magnon energies and exchange integrals, respectively, are smaller by
some orders of magnitude than other energies in the system as the d-f
exchange J or the conduction band width W. Accordingly, the respective
terms ($J^{\alpha\beta}_{ij},D_0$) can be neglected when dealing with
the electronic excitation spectra. That does not at all mean that there
is no influence of the magnetic moments on the electronic quasiparticle
spectrum. That is done by the spin correlation functions of the last
Section.

Starting from the Hamiltonian of the electronic subsystem,
\begin{equation}
\label{eq:h_*}
{\cal H}^* = {\cal H}_d + {\cal H}_{df},
\end{equation}
all physical relevant information of the system can be derived from the
retarded single-electron Green function:
\begin{eqnarray}
  \label{eq:green_gen}
  \lefteqn{G^{mm'}_{ij\alpha\beta\sigma} (E)  =  \left\langle\!\left\langle 
  c_{i\alpha m\sigma};
  c^+_{j\beta m'\sigma} \right\rangle\!\right\rangle_E \nonumber}
\hspace{1em} \\ 
 & = & - {\rm i} \int\limits_0^{\infty} dt \:
  {\rm e}^{-\frac{{\rm i}}{\hslash} Et} \left\langle \left[ 
  c_{i\alpha m\sigma}(t), c^+_{j\beta m'\sigma}(0) \right]_+ \right\rangle.
\end{eqnarray}
Here and in what follows $[.,.]_+$ ($[.,.]_-$) is the anticommutator
(commutator). Conform to the two-dimensional translational symmetry, we
perform a Fourier transformation within the layers of the film,
\begin{equation}
  \label{eq:green_gen_k}
  G^{mm'}_{{\bf k}\alpha\beta\sigma}(E) = \frac{1}{N} \sum_{ij} 
  {\rm e}^{{\rm i}{\bf k}({\bf R}_i - {\bf R}_j)} 
  G^{mm'}_{ij\alpha\beta\sigma}(E),
\end{equation}
where $N$ is the number of sites per layer, ${\bf k}$ is an in-plane
wavevector from the first 2D-Brillouin zone of the layers and ${\bf
  R}_i$ represents the in-plane part of the position vector, \mbox{${\bf
    R}_{i\alpha} = {\bf R}_i + {\bf r}_{\alpha}$}. From Eq.\ 
(\ref{eq:green_gen_k}) we get the spectral density by
\begin{equation}
  \label{eq:spectral}
  S^{mm'}_{{\bf k}\alpha\beta\sigma}(E) = - \frac{1}{\pi} {\rm Im} 
  G^{mm'}_{{\bf k}\alpha\beta\sigma} (E+{\rm i}0^+),
\end{equation}
which is directly related to observable quantities within angle and spin
resolved direct and inverse photoemission experiments. Finally, the
wave-vector summation of $S^{mm}_{{\bf k}\sigma}(E)$ yields the
layer-dependent (local) quasiparticle density of states:
\begin{equation}
\label{eq:ldos}
\rho^{m}_{\alpha\sigma}(E) = \frac{1}{\hslash N} \sum_{{\bf k}} 
S^{mm}_{{\bf k}\alpha\alpha\sigma}(E).
\end{equation} 
In the following discussion all results will be interpreted in terms of
the spectral density (\ref{eq:spectral}) and the local density of states
(\ref{eq:ldos}).

For the solution of the many-body problem posed by Eq.\ (\ref{eq:h_*})
we write down the equation of motion of the single-electron Green
function (\ref{eq:green_gen})
\begin{eqnarray}
  \label{eq:eom_gen}
  E\,G^{mm'}_{ij\alpha\beta\sigma} &=& \hslash\,\delta_{ij}
  \delta_{\alpha\beta} \delta_{mm'}\nonumber\\&& +
  \sum_{k\gamma m''} T^{mm''}_{ik\alpha\gamma}
  G^{m''m'}_{kj\gamma\beta\sigma}\nonumber\\&& +  
  \langle\!\langle [c_{i\alpha m\sigma},{\cal H}_{df}]_-;
  c^+_{j\beta m'\sigma} \rangle\!\rangle_E.
\end{eqnarray}
The formal solution of Eq.\ (\ref{eq:eom_gen}) can be found by
introducing the self-energy $M^{mm'}_{ij\alpha\beta\sigma}(E)$,
\begin{eqnarray}
  \label{eq:self_gen}
 \left\langle\!\left\langle \left[c_{i\alpha m\sigma},{\cal H}_{df}\right]_-;
  c^+_{j\beta m'\sigma} \right\rangle\!\right\rangle_E =\nonumber\\
  \sum_{k\gamma m''} M^{mm''}_{ik\alpha\gamma\sigma}(E)
  G^{m''m'}_{kj\gamma\beta\sigma}(E),   
\end{eqnarray}
which contains all information about the correlations between the
conduction band and localized moments. After combining Eqs.\ 
(\ref{eq:eom_gen}) and (\ref{eq:self_gen}) and performing a
two-dimensional Fourier transform we see that the formal solution of
Eq.\ (\ref{eq:eom_gen}) is given by
\begin{equation}
  \label{eq:form_sol}
  {\bf G}_{{\bf k}\sigma}(E) = \hslash \left(E\,{\bf I}-{\bf T}_{\bf k} -
  {\bf M}_{{\bf k}\sigma}(E) \right)^{-1}.
\end{equation}
Here, ${\bf I}$ represents the ($nM\times nM$) identity matrix, with $M$
denoting the number of subbands of the conduction band system, and the
matrices ${\bf G}_{{\bf k}\sigma}(E)$, ${\bf T}_{\bf k}$, and ${\bf
  M}_{{\bf k}\sigma}(E)$ have as elements the layer- and
subband-dependent functions $G^{mm'}_{{\bf k}\alpha\beta\sigma}(E)$,
$T^{mm'}_{{\bf k}\alpha\beta}$, and $M^{mm'}_{{\bf
    k}\alpha\beta\sigma}(E)$, respectively.

To explicitly get the self-energy in Eq.\ (\ref{eq:self_gen}) we
evaluate the Green function
\begin{eqnarray}
  \label{eq:c_hdf}
  \left\langle\!\left\langle \left[c_{i\alpha m\sigma},{\cal H}_{df}\right]_-;
  c^+_{j\beta m'\sigma} \right\rangle\!\right\rangle_E =\nonumber\\
  - \frac{J}{2} \left( z_{\sigma} 
  {\it \Gamma}^{mm'}_{iij\alpha\alpha\beta\sigma} 
  + F^{mm'}_{iij\alpha\alpha\beta\sigma} \right).
\end{eqnarray}
Here, the two higher Green functions,
\begin{eqnarray}
  \label{eq:gamma}
  {\it \Gamma}^{m''m'}_{ikj\alpha\gamma\beta\sigma}(E) & = &
  \left\langle\!\left\langle S^z_{i\alpha} c_{k\gamma m''\sigma} ;
  c^+_{j\beta m'\sigma} \right\rangle\!\right\rangle_E ,\\
  \label{eq:spin-flip}
  F^{m''m'}_{ikj\alpha\gamma\beta\sigma}(E) & = &
  \left\langle\!\left\langle S^{-\sigma}_{i\alpha} c_{k\gamma m''-\sigma} ;
    c^+_{j\beta m'\sigma} \right\rangle\!\right\rangle_E ,
\end{eqnarray}
originate form the two terms of the {\em d-f} Hamiltonian (\ref{eq:H_df}) and will
be referred to as the {\em Ising} and the {\em Spin-flip} function,
respectively. Considering the equations of motion for these two Green
functions we encounter the two higher Green functions $\langle\!\langle
[S^z_{i\alpha} c_{k\gamma m''\sigma}, {\cal H}_{df}]_- ; c^+_{j\beta
  m'\sigma} \rangle\!\rangle_E$ and $\langle\!\langle
[S^{-\sigma}_{i\alpha} c_{k\gamma m''-\sigma}, {\cal H}_{df} ]_- ;
c^+_{j\beta m'\sigma} \rangle\!\rangle_E$.  Since we consider an empty
conduction band the thermodynamic average in the Green functions has to
be computed with the electron vacuum state $\left| n=0 \right\rangle$.
From the definition of the {\em d-f} Hamiltonian (\ref{eq:H_df}) we then see that
$\left\langle n=0\right|{\cal H}_{df}=0$ and, accordingly,
\begin{eqnarray*}
  \left\langle\!\left\langle \left[S^z_{i\alpha},{\cal H}_{df}\right]_-
  c_{k\gamma m''\sigma};c^+_{j\beta m'\sigma} \right\rangle\!\right\rangle_E
  & \stackrel{n\rightarrow 0}{\longrightarrow} & 0 , \\
  \left\langle\!\left\langle \left[S^{-\sigma}_{i\alpha},{\cal H}_{df}\right]_-
  c_{k\gamma m''-\sigma};c^+_{j\beta m'\sigma} \right\rangle\!\right\rangle_E
  & \stackrel{n\rightarrow 0}{\longrightarrow} & 0 .
\end{eqnarray*}
Hence, for the equations of motion of the Ising and the Spin-flip
function we get
\begin{eqnarray}
  \label{eq:eom_ga_gen}
  \lefteqn{\sum_{l\delta m'''} \left(E\delta_{kl} \delta_{\gamma\delta}
      \delta_{m''m'''}-T^{m''m'''}_{kl\gamma\delta} \right) 
  {\it \Gamma}^{m'''m'}_{ilj\alpha\delta\beta\sigma}(E)} \nonumber \\ & = & 
  \hslash \left\langle S^z_{\alpha} \right\rangle
  \delta_{kj} \delta_{\gamma\beta} \delta_{m''m'}+\nonumber\\
 && + \left\langle\!\left\langle S^z_{i\alpha} \left[ c_{k\gamma m''\sigma},
  {\cal H}_{df} \right]_- ; c^+_{j\beta m'\sigma}
  \right\rangle\!\right\rangle_E \\[2ex]
  \lefteqn{\label{eq:eom_f_gen}
  \sum_{l\delta m'''} \left(E\delta_{kl} \delta_{\gamma\delta}
    \delta_{m''m'''} - T^{m''m'''}_{kl\gamma\delta} \right) 
  F^{m'''m'}_{ilj\alpha\delta\beta\sigma}(E)} \nonumber \\ & = &
  \left\langle\!\left\langle S^{-\sigma}_{i\alpha} 
  \left[ c_{k\gamma m''-\sigma}, {\cal H}_{df} \right]_- ; 
  c^+_{j\beta m'\sigma} \right\rangle\!\right\rangle_E .
\end{eqnarray}
On the right-hand side of these equations appear further higher Green
functions which prevent a direct solution and require an approximative
treatment. The treatment is different for the non-diagonal terms,
$(i,\alpha,m)\ne(k,\gamma,m'')$ and for the diagonal terms,
$(i,\alpha,m)=(k,\gamma,m'')$. In the first case we use a
self-consistent so-called {\em self-energy approach} which results in a
decoupling of the equations of motion.  For the diagonal terms,
$(i,\alpha)=(k,\gamma)$, this approach is replaced by a moment technique
which takes the local correlations better into account.

The details of the approach have already been published in our previous
paper \cite{SN99}. By introducing the multi-indices $A,B,\ldots $,
defined by$(\alpha ,m)\longrightarrow A$, $(\beta
,m^{\prime})\longrightarrow B,\ldots$, the approximative solution for
the electronic subsystem of the single-band KLM presented in \cite{SN99}
can be transferred to the case of the multi-band KLM, with the
correspondence $\alpha\cong A,\beta\cong B,\ldots$.

The reason for this straightforward transference is that the commutator
relations which govern the solution in \cite{SN99} are invariant under
the index transition,e.g.
\begin{eqnarray}
\left[ c_{i\alpha\sigma},H_{sf}^{sb}\right]_{-}&=&- \frac{J}{2} (z_{\sigma}\,{\bf
  S}_{i\alpha}^{z}\,c_{i\alpha\sigma}+{\bf
  S}^{-\sigma}_{i\alpha}\,c_{i\alpha -\sigma})\nonumber\\
\left[ c_{iA\sigma},H_{df}^{mb}\right]_{-}&=& -\frac{J}{2} (z_{\sigma}\,{\bf
  S}_{i\alpha}^{z}\,c_{i\alpha m\sigma}+{\bf S}^{-\sigma}_{i\alpha}\,c_{i\alpha m -\sigma})\nonumber\\
&=&-\frac{J}{2}(z_{\sigma}\,{\bf S}_{iA}^{z}\,c_{iA\sigma}+{\bf S}^{-\sigma}_{iA}\,c_{iA -\sigma})
\end{eqnarray}
Thus there is no need to repeat the procedure once more. For details the
reader is referred to \cite{SN99}
\section{Results for EuO-films}\label{sec:results-euo-films}
EuO crystallizes in the NaCl structure. Hence, the $Eu^{2+}$ ions are
arranged on an fcc lattice. For all the following film calculations, we
assume the surface to be parallel to the fcc(100) crystal plane.
\subsection{Magnetic properties}\label{sec:magnetic-properties}
The calculation of the magnetic properties of the EuO films has been
done with the theory of Sect.III.A. We take the Heisenberg-exchange
integrals from the experiment. Low temperature neutron scattering
measuring the spin-wave dispersion \cite{BZDK80} and being compared to
the renormalized spin wave theory \cite{quantmag2} have revealed that
nearest ($J_1$) and next-nearest neighbor ($J_2$) exchange integrals
have to be taken into account:
\begin{eqnarray}
\frac{J_1}{k_B}&=&0.625\,K\nonumber\\
\frac{J_2}{k_B}&=&0.125\,K
\end{eqnarray}
For an fcc (100) film, we have therefore for the case of uniform $J_1$
and $J_2$ within the film:
\begin{eqnarray}
J^{\alpha\beta}_{ij}&=&J_1
({\delta}^{\alpha\beta}_{i,j+\Delta_{1\|}}+{\delta}^{\alpha\beta\pm
  1}_{i,j+\Delta_{1\bot}})\nonumber\\
&&+J_2({\delta}^{\alpha\beta}_{i,j+\Delta_{2\|}}+{\delta}^{\alpha\beta\pm
  2}_{i,j})
\end{eqnarray}
Here,$\Delta_{1\|}$,$\Delta_{1\bot}$ and $\Delta_{2\|}$
denote,respectively, the positions of nearest neighbors within the same
layer, within the adjacent layers and of next-nearest neighbors within
the same layer one finds,
\begin{eqnarray}\label{eq:delta_all}
2\Delta_{1\|}&=&(1,1),(1,-1),(-1,1),(-1,-1)\nonumber\\
2\Delta_{1\bot}&=&(0,1),(0,-1),(1,0),(-1,0)\nonumber\\
\Delta_{2\|}&=&(0,1),(0,-1),(1,0),(-1,0)\nonumber\\
\Delta_{2\bot}&=&(0,0)
\end{eqnarray}
\begin{figure}[t]
  \centerline{\epsfig{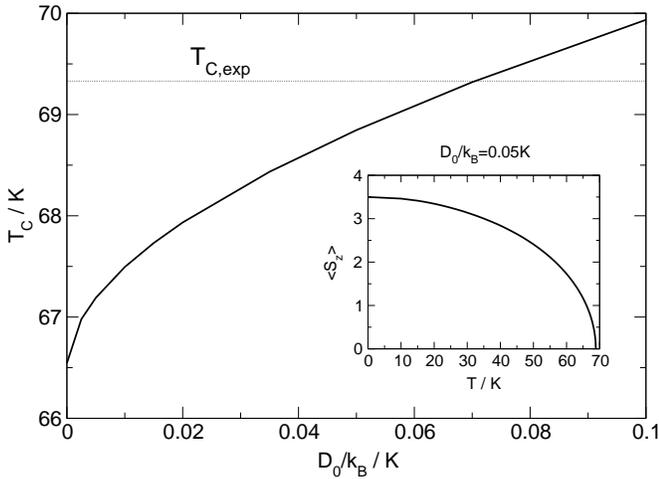}}
    \caption{Dependence of the Curie temperature $T_{\mathrm{C}}$ on the single-ion
      anisotropy constant $D_0$, calculated for bulk EuO. Dotted line
      represents the experimental value, ${T_{\mathrm{C}}}=69,33K$. Inset:
      Magnetization of bulk EuO as function of temperature, calculated
      for ${D_0}=0.005k_B$}
    \label{fig:Nol:fig5}
\end{figure}
\begin{figure}[t]
  \centerline{\epsfig{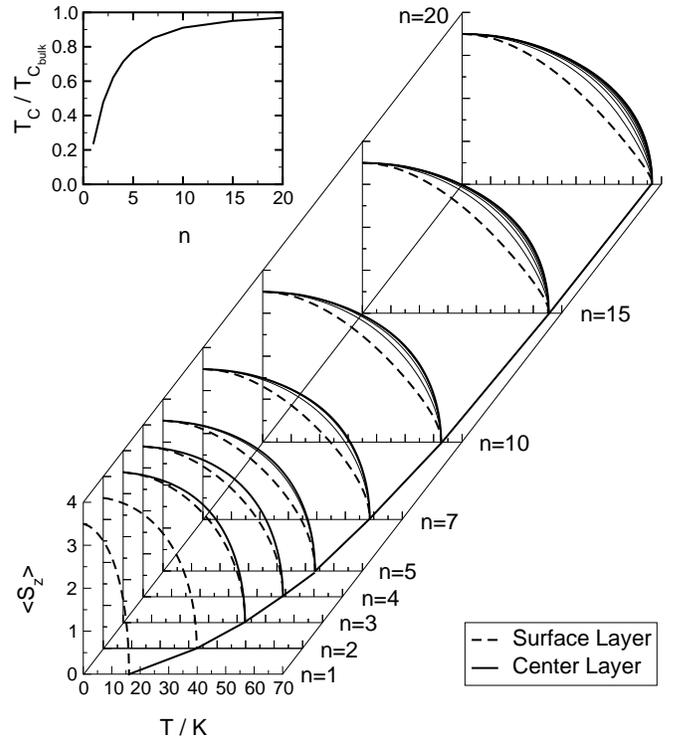}}
    \caption{Layer-dependent magnetizations, $\langle {\bf
        S}^{z}_{\alpha} \rangle$, of EuO(100) films as functions of
      temperature for various thicknesses n. $\langle {\bf S}^{\alpha}
      \rangle$ always increases monotonously from the surface layer
      towards the film center. Inset: Curie temperature as a function of
      film thickness.}
    \label{fig:Nol:fig6}
\end{figure}
It remains to fix in a reasonable manner the single-ion anisotropy
constant $D_0$. For this purpose we have calculated the Curie
temperature of bulk (!) EuO with $J_1$,$J_2$ from (\ref{eq:delta_all}) and variable
$D_0$. The result is shown in Fig.~\ref{fig:Nol:fig5}. When
$\frac{D_0}{k_B}$ increases from 0 to 0.1K $T_{\mathrm{C}}$ rises from about 66.5K
to 70K. Regarding that $J_1$,$J_2$ stem from a low-temperature fit, the
agreement between the calculated $T_{\mathrm{C}}$'s and the experimental value of
69.33K \cite{Wac79} is remarkably good and that for all $D_0$-values in
the investigated region. We have chosen $\frac{D_0}{k_B}=0.05K$ for our
calculations of film geometries. This is in reasonable agreement with
the rather spare experimental values. According to
(\ref{eq:sol_eom_heis}) and (\ref{eq:a_al_be}) the
model anisotropy energy is $E_a = -D_0\,\Phi_{\alpha}$ with the
temperature-dependent anisotropy coefficient $\Phi_{\alpha}$ from Eq.
(\ref{eq:lines}). It can be seen that $E_a$ monotonically decreases as
function of 
$\langle S^z_{\alpha}\rangle$ to $E_a=-6D_0$ at $\langle
S^z_{\alpha}\rangle=S$. That means at $T=0$ we have
$\frac{E_a}{k_B}\approx -0.3$ which is the same order of magnitude as
the experimental value in ref.\cite{MA67}.

Fig.~\ref{fig:Nol:fig6} shows the temperature and layer-dependent
magnetizations of EuO(100) films for thicknesses from $n=1$ to $n=20$.
For all temperatures and for all film thicknesses the $\langle
S^z_{\alpha}\rangle$ increase from the lowest magnetization of the
surfaces monotonously towards the highest magnetizations in the film
centers. That can qualitatively be explained by the lower coordination
number of the surface atoms, $Z_{S,fcc(100)}=4$, compared to the bulk
coordination number, $Z_{b,fcc}=12$. The inset of
Fig.~\ref{fig:Nol:fig6} shows the dependence of the Curie temperature on
the film thickness n compared to the bulk-$T_{\mathrm{C}}$ calculated with the same
parameters $J_1$,$J_2$, and $D_0$ (see inset of
Fig.~\ref{fig:Nol:fig5}). Starting from a rather low value for the
monolayer $T_{\mathrm{C}}$ steadily increases with n reaching the bulk-value for
$n\geq 20$. The curve strongly resembles that of Gd films \cite{Far93}.
Corresponding measurements for EuO are unknown to us.

All the higher spin correlation functions $\langle
S^{\pm}_{\alpha}S^{\mp}_{\alpha}\rangle$, $\langle (S^{z}_{\alpha})^2
\rangle$, $\langle(S^{z}_{\alpha})^3 \rangle$, $\ldots$ follow from
$\langle S^{z}_{\alpha} \rangle$ via (\ref{eq:s_z}),
(\ref{eq:rel_coeff}), (\ref{eq:spin_op_rel}) . Examples
are plotted in Fig.~\ref{fig:Nol:fig2} of ref.\cite{SN99}. Together with
$\langle S^{z}_{\alpha} \rangle$ they provide the necessary temperature
information needed to calculate the temperature-dependent electronic
structures of EuO films, which are presented in the next Section.
\subsection{Temperature-dependent electronic structure}\label{sec:temp-depend-electr}
In order to get the temperature-dependent band structures of the
EuO(100) films we have to combine the LSDA calculations, described in
Sect.II.B, with the many-body evaluation of the multiband-Kondo lattice
model (Sect.III.B). It has been shown in previous works
\cite{NMR96,NMR97,SN99} that the d-f exchange interaction of the KLM can
be treated exactly for T=0 in the case of empty conduction bands. While
the $ \downarrow $-spectrum turns out to be rather complicated
exhibiting interesting correlation effects, the $\uparrow$-spectrum is
simple being only slightly shifted towards lower energies by a constant
energy amount of $\frac{1}{2}J\, S$. So we can directly use the TB-LMTO
obtained ${\bf k}$-dependent hopping matrices for the
$\uparrow$-electron as input for the single-particle Hamiltonian $H_d$
(\ref{eq:h_d}). The problem of double-counting of relevant interactions, here the
d-f-exchange interaction, which usually occurs when combining
first-principles and model calculations, is then elegantly avoided.
\begin{figure}[b]
  \centerline{\epsfig{file=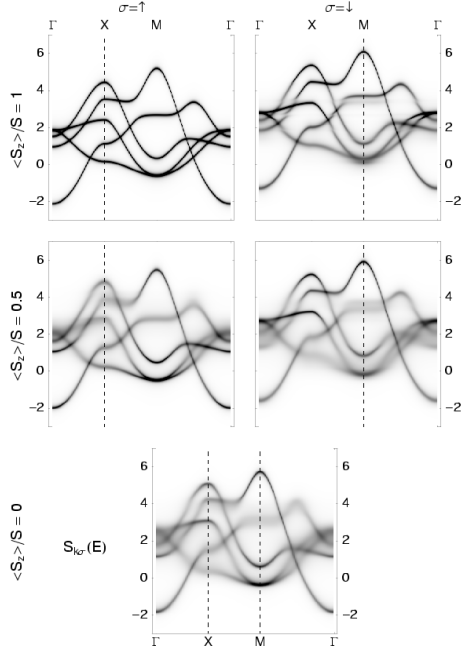,width=\linewidth}}
    \caption{Spin-dependent spectral densities of the Eu-5d bands of a
      EuO(100)monolayer for different magnetizations $\langle \frac{{\bf
          S}^z}{{\bf S}} \rangle$ (see Table \ref{tab:sz-t}).}
    \label{fig:Nol:fig7}
\end{figure}
The temperature comes into play via the layer- and temperature-dependent
f-spin correlation functions, discussed in the last section, which
appear in the many-body evaluation of the KLM (Sect.~\ref{sec:conduction-bands}). To avoid
ambiguities, we use in the following as temperature-parameter the
reduced magnetization $\frac{\langle S^{z}\rangle}{S}$ of the center
layer of the respective EuO (100) film. The same temperature may produce
in the other layers different magnetizations.

\begin{table}[t!]
  \centerline{\ttfamily
    \begin{tabular}{|l|l|l|l|l|l|l|}
      \hline
      &\multicolumn{1}{c|}{\raisebox{-.4ex}[.4ex]{\rm bulk}} 
      &\multicolumn{5}{|c|}{\rm EuO(100) films (number of layers)}\\ 
      \cline{3-7} 
      \raisebox{1.5ex}[-1.5ex]{$\langle S_z \rangle/S$}
      &\multicolumn{1}{c|}{\rm EuO}
      &\multicolumn{1}{c|}{\rm 1}
      &\multicolumn{1}{c|}{\rm 2}
      &\multicolumn{1}{c|}{\rm 5}
      &\multicolumn{1}{c|}{\rm 10}
      &\multicolumn{1}{c|}{\rm 20}
      \\ \hline\hline 
      1 & ~0 & ~0 & ~0 & ~0 & ~0 & ~0\\[-0.8ex] 
      0.75 & 45.48 & 10.52 & 21.56 & 37.41 & 43.23 & 45.14\\[-0.8ex]
      0.5 & 59.76 & 13.95 & 28.51 & 47.51 & 55.59 & 58.92\\[-0.8ex]
      0.25 & 66.71 & 15.56 & 31.82 & 52.02 & 61.02 & 65.04\\[-0.8ex]
      0 \scriptsize$(T_{\rm C})$& 68.84 & 16.06 & 32.83 & 53.36 & 62.70 & 66.73\\ \hline
    \end{tabular}
    }
    \caption{Temperatures at which the 4f magnetization of bulk EuO 
      and of the center layers of EuO (100) films has assumed the
      values in the left column (see Figs.~\ref{fig:Nol:fig5}, \ref{fig:Nol:fig6})
    \label{tab:sz-t}}
\end{table}
Table \ref{tab:sz-t} gives some examples how the different values of the
magnetization $\frac{\langle S^{z}\rangle}{S}$ are related to the
respective temperatures for EuO (100) films.
\begin{figure}[t]
  \centerline{\epsfig{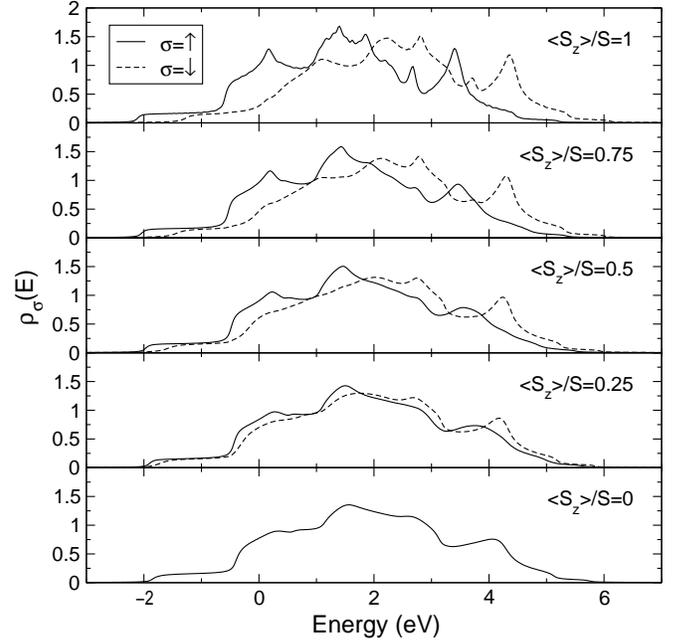}}
    \caption{Temperature-dependent densities of states of the Eu-5d
      bands of a EuO(100) monolayer for different values of the 4f
      magnetization $\langle \frac{{\bf S}^z}{{\bf S}} \rangle$ (see
      Table \ref{tab:sz-t}). }
    \label{fig:Nol:fig8}
\end{figure}
The only parameter of our theory, the exchange coupling J, is chosen
from bulk-EuO calculations \cite{SN01a} to yield the experimentally
observed `red shift` of the lower conduction band edge upon cooling down
from $T=T_{\mathrm{C}}$ to $T=0$:
\begin{equation}
J=0.25\,eV
\end{equation}
\begin{figure}[t]
  \centerline{\epsfig{file=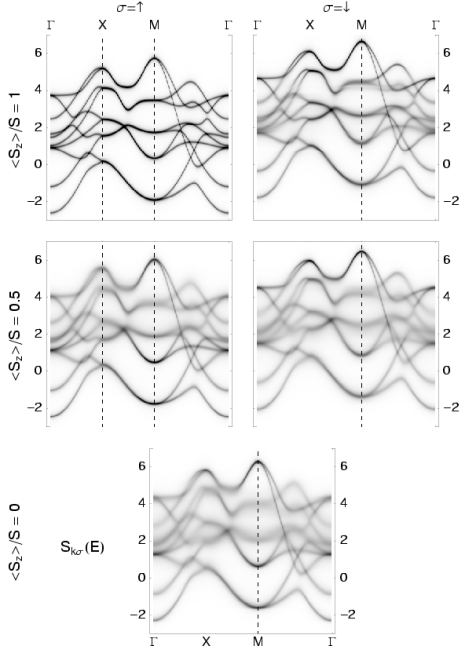,width=\linewidth}}
    \caption{Spin-dependent spectral densities of the Eu-5d bands of an
      EuO(100) double layer (n=0) for different values of the 4f
      magnetization $\langle \frac{{\bf S}^z}{{\bf S}} \rangle$ (see
      Table \ref{tab:sz-t}).}
    \label{fig:Nol:fig9}
\end{figure}
\begin{figure}[b]
  \centerline{\epsfig{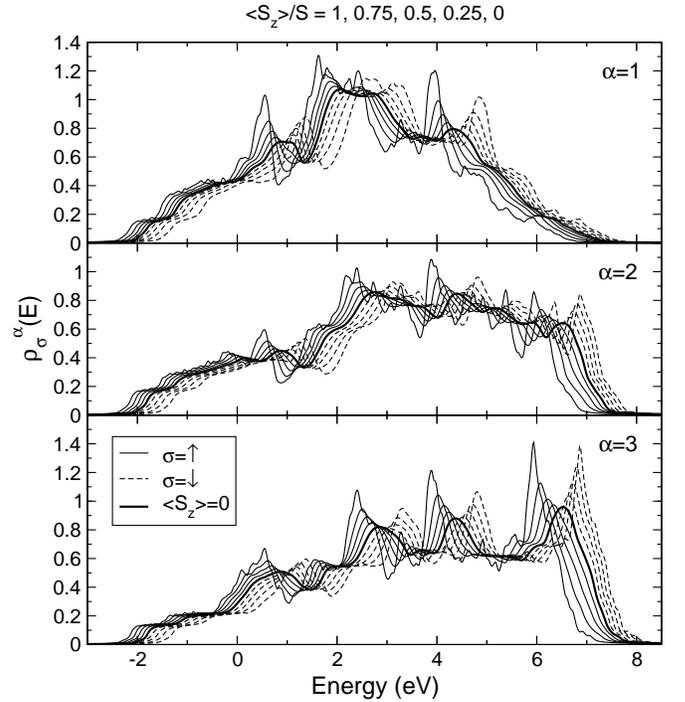}}
    \caption{Local densities of states of the Eu-5d bands of the first,
      second and center layer of a 5-layer EuO(100) film for different
      values of the center layer 4f magnetization $\frac{\langle
        S^{z}\rangle}{S}$ (see Table \ref{tab:sz-t}).}
    \label{fig:Nol:fig10}
\end{figure}
\begin{figure}[t]
  \centerline{\epsfig{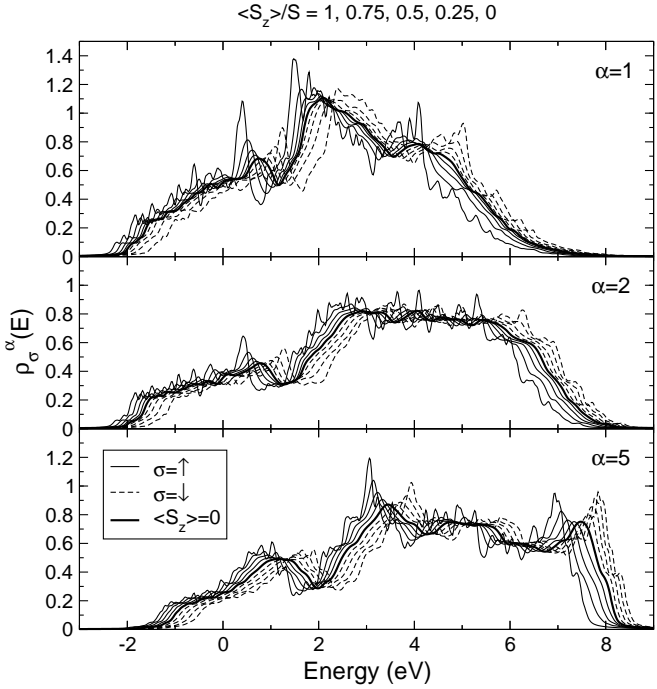}}
    \caption{The same as in Fig.~\ref{fig:Nol:fig10} but for a 10-layer film}
    \label{fig:Nol:fig11}
\end{figure}
Let us start the discussion with the spectral densities and the
quasiparticle densities of states (Q-DOS) of the Eu-5d bands for a
EuO(100) monolayer (n=1) at varying magnetizations $\frac{\langle
  S^{z}\rangle}{S}$ of the 4f moment system. Fig.~\ref{fig:Nol:fig7}
shows the quasiparticle band structure as a density plot. The degree of
blackening is a measure of the peak-height of the respective spectral
density. Broad, washed out lines mean quasiparticles with relatively low
life-times. In ferromagnetic saturation ($\frac{\langle S^{z}\rangle}{S}
=1$) the $\uparrow$-spectrum consists of stable quasiparticles. As
explained above this is the exactly solvable limiting case. The spectrum
is identical to the LSDA input only slightly shifted by $-\frac{1}{2}J\,
S$. The $\downarrow$-spectrum is also exactly solvable exhibiting,
however, clear correlation effects as, e.g., a splitting of the
dispersion around 4eV near the M-points. As discussed in detail for a
model system in ref.\cite{NMR96,SN99} the two excitations belong to two
different microscopic interaction processes. The high-energy part is due
to the formation of a magnetic polaron (propagating electron `dressed`
by a virtual cloud of magnons), the low-energy branch is due to
scattering processes (magnon-emission by the $\downarrow$-electron). The
magnetic polaron can be considered as due to a polarization of the
neighboring 4f spins by the conduction electron via d-f exchange. Under
certain parameter constellations this may even lead to a bound
state,e.g. to a quasiparticle with infinite lifetime \cite{NMR96,SN99}.
This exchange-induced splitting, which can also be observed in other
parts of the T=0-$\downarrow$-band structure, is a typical correlation
effect, which by no means can be accounted for in an LSDA calculation.

An exchange splitting of the full $\uparrow$- and $\downarrow$ spectra
against one another is clearly visible. Note that these effects appear
although the bands are empty, more strictly there is one electron (`test
electron`) in the otherwise empty 5d-bands. They are therefore certainly
induced by the 4f system and mediated by the d-f exchange interaction.

For finite temperature, intermediate magnetizations,$\frac{\langle
  S^{z}\rangle}{S}=0.5$, the $\uparrow$-dispersion, too, starts to wash
out indicating spin-flip processes. Magnon emission by a
$\downarrow$-electron is completely equivalent to magnon absorption by a
$\uparrow$-electron, with one exception: magnon absorption is possible
only if there are any in the spin system. This is not the case in
ferromagnetic saturation ($\langle S^{z}\rangle=S$). That is the reason
why at T=0K the $\uparrow$-spectrum appears so much simpler than the
$\downarrow$-spectrum. For finite temperatures (demagnetizations)
magnons are available, absorption processes lead to damping of the
quasiparticle dispersion in the $\uparrow$-spectrum, too. The overall
exchange splitting reduces with increasing temperatures and the two spin
spectra are becoming more and more similar. Finally, in the limit of
$\langle S^{z}\rangle\longrightarrow 0$, $T\longrightarrow T_{\mathrm{C}}$ the lack
of any 4f-magnetization removes the induced spin-asymmetry in the 5d
bands (Fig.~\ref{fig:Nol:fig7}).

The temperature-dependent Q-DOS of the EuO-5d bands in a monolayer
exhibits a strong induced exchange splitting at T=0
(Fig.~\ref{fig:Nol:fig8}), where the increase of temperature from T=0 to
$T=T_{\mathrm{C}}$ leads to a convergence of the two spin parts. It can clearly be
seen that the temperature dependence of the exchange-split spectra goes
beyond the Stoner picture, where the bands move rigidly towards each
other when the temperature is increased. The reason lies predominantly
in the spin-flip processes between 5d- and 4f- system. Such correlation
effects take care for the non-rigid shift. By comparison with the
$\uparrow$-Q-DOS for ferromagnetic saturation ($\langle
S^{z}\rangle=S$), the shape of which is identical to the
LDA-renormalized single-particle input, one recognizes that in the
paramagnetic phase, too, correlation effects remarkably alter the
spectrum.

Fig.~\ref{fig:Nol:fig9} displays the spectral density of a EuO double
layer for three different magnetizations of the 4f-moment system. The
same which has been said above about the temperature-dependent
tendencies of the monolayer EuO(100) spectra applies to the spectra of
the double-layer in Fig.~\ref{fig:Nol:fig9}. However, due to the higher
number of bands which originates from the hybridization of the two
EuO(100) layers the temperature-dependent evolution of the bands becomes
less clear for the EuO(100) double layer.

Figs.~\ref{fig:Nol:fig10} and \ref{fig:Nol:fig11} show the
temperature-dependent Q-DOS of, respectively, a 5-layer and a 10-layer
EuO(100) film. Different from the case of the monolayer and the double
layer, for these films the densities of states depend on the layer index
$\alpha$. Furthermore, the magnetization of the 4f-moment system is a
layer-dependent entity (see Fig.~\ref{fig:Nol:fig6}). Here, the
temperature parameter $\frac{\langle S^{z}\rangle}{S}$ refers to the
magnetization of the center layer of the film. The respective
temperatures are given in Table \ref{tab:sz-t}.

With increasing film thickness, the Q-DOS of the center layer becomes
more and more similar to the Q-DOS of bulk EuO. As a result, the centers
of gravity of the surface layer bands, on the one hand, and of the
center layers, on the other hand, move away from each other, with the
latter positioning at higher energies. This tendency leads to the
existence of EuO(100) surface states, which are discussed in detail in
an additional paper \cite{SN01}.

There it is shown that for $T=T_{\mathrm{C}}$ the surface at the
$\Gamma$--point lies some 0.8 eV below the bulk conduction band edge.
This distance does not change with temperature, t.e.\ the $\uparrow$
part of the surface state exhibits the same red shift of about 0.3 eV as
the 5d conduction band edge when cooling down to
$T_{\mathrm{C}}=0\mathrm{K}$. Since the paramagnetic 4f--5d gap amounts
to 1.12 eV\cite{Wac79} it can be speculated\cite{SN01} that the
$\uparrow$--surface state closes the gap at very low temperatures.
Giving rise to a temperature--induced surface insulator--halfmetal
transition. The Eu--ion then remains in an intermediate valence
fluctuating between the magnetic $4\mathrm{f}^7$--state
($J=S=\frac{7}{2}$) and the non-magnetic $4\mathrm{f}^6$--state
($J=0$). Interesting dynamic effects are to be expected since the
non-magnetic $4\mathrm{f}^6$--state automatically cancels the red shift
pushing the $\uparrow$--surface states by 0.3 eV to higher
energies. Electrons which will reenter the 4f level creating again the
magnetic  $4\mathrm{f}^7$ configuration, and so on.

\section{Conclusions}\label{sec:conclusions}
By combination of a `first-principles` band structure calculation within
the frame of density functional theory with a many-body evaluation of a
proper theoretical model we have derived the temperature-dependent
electronic structure and the magnetic properties of a real 4f system
(EuO) with film geometry. The theory provides the properties of films of
varying thicknesses. The film geometry provokes a superposition of
electron correlation effects with the sheer geometrical dependence of
the spectra on film thickness and layer index.

`Local-moment` systems like EuO get their striking temperature
dependencies and correlation effects by an exchange interaction between
localized magnetic moments, due to 4f electrons of the Rare Earth ion,
and quasi-free (5d,6s) conduction electrons. The situation is rather
convincingly modeled by the Kondo-lattice (s-f, s-d) model (KLM). We
proposed a way how to generalize the original single-band KLM to a
realistic multi-band model. The respective many-body problem is
approximately solved for the electronic subsystem making use of a
moment-conserving decoupling approximation (MCDA) for suitably defined
Green functions. The method gains substance by the fact that the
finite-temperature theory evolves continuously from the exactly solvable
and nontrivial T=0 case of ferromagnetic saturation.

The mentioned T=0-limiting case is also used to connect the
multiband-model treatment with a TB-LMTO-ASA band structure calculation
for thin EuO(100) films. Circumventing the much discussed
double-counting problem the spin-up Eu-5d bands have been taken as input
for the hopping matrix of the multi-band-KLM Hamiltonian. The reason is
that the exact T=0-KLM solution shows for ferromagnetic saturation that
the $ \uparrow $-quasiparticle spectrum is only rigidly shifted compared
to the `free` dispersions. A double counting of the d-f exchange
interaction is therefore excluded.

The film geometries are accounted for by a supercell construction, where
consecutive EuO(100) films are isolated from each other by a stack of
spacer layers. The connection to the model calculation employs a
decomposition into subbands, that respects the symmetry of the Eu-5d
orbitals.

Already for the case of ferromagnetic saturation of the 4f-moment
system, the spin-down spectra of the EuO films exhibit strong deviations
from the single-electron LSDA results demonstrating strong correlation
effects. These correlation effects in the $\downarrow$-spectra provoke a
strong temperature-dependence. The $\uparrow$-spectra, which at T=0
correspond to the $\uparrow$-LSDA band structure results, develop
similar correlation effects for finite temperatures. The loss of 4f
magnetization for $T\longrightarrow T_{\mathrm{C}}$ leads to a convergence of the
spectra for both spin directions.

The semiconductor EuO fits the physics of the KLM for the special case
of an empty conduction band. The extension to non-zero band occupation
allows to study the prototypical `local moment` metal Gd. Concerning the
aspect of reduced dimensionality, the dependence of magnetic and
electronic properties of Gd films are of high experimental and
theoretical interest. Respective forthcoming investigations will shed
light on some extraordinary physical properties of the lanthanides, in
particular the enhanced Curie temperature of the Gd(0001) surface and
its relation to the temperature behavior of the Gd(0001) surface state.
\subsection*{Acknowledgment}
Financial support by the `Sonderforschungsbereich 290` and by the German
Merit Foundation is gratefully acknowledged.
%
%
%\section*{References}

%

%
%
\end{document}